# Probing Brain Oxygenation with Near Infrared Spectroscopy (NIRS) — The Role of Carbon Dioxide and Blood Pressure

Alexander Gersten

Additional information is available at the end of the chapter

## 1. Introduction

The Near Infrared spectroscopy (NIRS) is a technique, which allows measuring the oxygenation of the brain tissue [22, 1, 6, 9, 37, 52, 58, 63].

The problem is to penetrate the brain with a light, which can cross the skin and skull and be absorbed mostly by the hemoglobin ($HbO_2$) and the deoxyhemoglobin (Hb). The ratio of the oxygenated hemoglobin ($HbO_2$) to the total hemoglobin (tHb=$HbO_2$+Hb) is the regional oxygen saturation rSO2. It appears that the most suitable light is in the 650-1000 nm range of the near infrared light. In this range, there exists another absorber cytochrome oxidase (CtOx), but as its concentration is quite small, it is often neglected [60]. Instruments that take into account the CtOx use 3 different wavelengths, otherwise 2 wavelengths are used.

There is a continuous improvement in the instrumentation. Three types are in use: continuous wave (CW) mostly used, time resolved and intensity modulated [9]. An improved resolution is obtained using the spatially resolved spectroscopy (SRS), in which multi-distance sensors are used. Another distinction can be made. Photometers, are the simplest, they use CW light, single-distance and one sensor. The oximeters use multi-distance (SRS) techniques with CW and usually two sensors [15, 9, 58]. Multi-channel CW imaging systems generating images of a larger area started to be used [15, 52].

We have done measurements with two instruments: the INVOS Cerebral Oximeter of Somanetics (www.somanetics.com; Thavasothy et al., 2002), and the hemoencephalograph (HEG) photometer of Hershel [66]. The results are presented in the following sections.

## 2. Theoretical considerations

The theoretical considerations are based on the modified Beer-Lambert law [22],

$$A = \left[ a_1 \cdot c_1 + a_2 \cdot c_2 + ... + a_n \cdot c_n \right] \cdot d \cdot DPF + G, \tag{1}$$

where $A$ is the attenuation coefficient, measured in optical densities (OD),

$$A = log_{10} \left[ I_0 / I \right] \tag{2}$$

$I_0$ is the incident light intensity and $I$ the intensity of the light collected by the sensor after returning from the brain.

It is a generalization of Beer-Lambert Law, in which the attenuation $A$ of an incident light is proportional to the concentration $c$ of the compound in the solution and the optical pathlength $d$:

$$A = log_{10} \left[ I_0 / I \right] = a \cdot c \cdot d, \tag{3}$$

a is the specific extinction coefficient of the absorbing compound (measured in 1/micromolar per cm), c is the concentration of the absorbing compound in the solution (measured in micromolar), $d$ is the distance in cm that the light covers in the medium, $a \bullet c$ is the absorption coefficient of the medium. For different absorbing compounds, the absorbing coefficient is the sum of the contributions,

$$A = log_{10} \left( I_0 / I \right) = \left[ a_1 \cdot c_1 + a_2 \cdot c_2 ... + a_n \cdot c_n \right] d. \tag{4}$$

The Beer-Lambert law should be modified to include a term $G$ that describes the scattering attenuation and a multiplying factor due to scattering, the differential pathlength factor (DPF), which multiplies the length d. All these corrections appear in the modified Beer-Lambert law as given in Eq. (1).In Fig.1 the specific extinction coefficients of hemoglobin (HbO$_2$), the deoxyhemoglobin (Hb) and cytochrome oxidase (CtOx) are displayed.

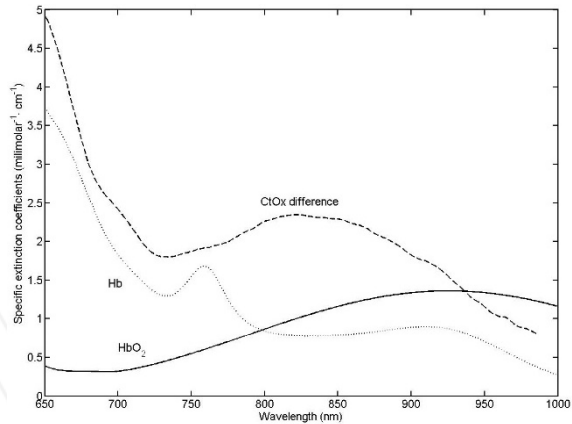

**Figure 1.** The specific extinction coefficients of HbO$_2$, Hb, and the difference absorption spectrum between the oxidized and reduced forms of CtOx, the data are from [7].

It appears that DPF depends on the human age. According to [7]:

$$DPF_{780} = 5.13 + \left(0.07Y\right)^{0.81},$$ (5)

where DPF$_{780}$ is the PDF at 780 nm and $Y$ is the age of the subject in years.

## 3. Oxygen utilization

For a more complete description of oxygen utilization see [22]. Here we will only mention [22] that "a typical oxygen carrying capacity of the blood is 19.4 ml of O$_2$ per dl of blood with 19.1 ml O$_2$/dl carried by hemoglobin and only 0.3 ml O$_2$/dl dissolved in plasma [7]. A typical averaged value for adult cerebral blood flow (CBF) is 47.7 ml/100 ml/min (Frackowiak et al., 1980) corresponding to total oxygen delivery of 9.25 ml O2/100 ml/min. [7]. Typical oxygen consumption of the adult brain is 4.2 ml O$_2$/100 ml/min (Frackowiak et al., 1980). CBF, cerebral blood volume (CBV) and cerebral oxygen extraction (COE) are significantly greater in grey matter compared to white matter in normal human adults (Lammertsma et al., 1983; cope, 1991). Only part of the arterial oxygen that arrives in the brain is absorbed and utilized. The fraction that is utilized, known as the oxygen extraction fraction (OEF), is defined as

$$OEF = \left(SaO_2 - SvO_2\right)/SaO_2,$$ (6)

where SaO$_2$ and SvO$_2$ are the arterial and venous oxygen saturations respectively."

In a well-done experiment, the EOF, measured in normal control subjects, was 0.41±0.03 [11]. Assuming $SaO_2$ equal to 0.95 (an average adult value) the $SvO_2$ will be, using Eq. (6), equal to 0.56±0.03. In the brain tissue, the absorption of the hemoglobin is mostly venous. A 75% venous contribution, leads in the frontal region to $rSO_2$ of about 66±3%, which is observed in experiments.

## 4. The INVOS oximeter

The Somanetics INVOS Cerebral Oximeter (ICO) uses two wavelengths, 730 and 810 nm and measures $rSO_2$ of the brain in the area under the sensor, which is attached to the forehead. The spatially resolved spectroscopy (SRS) method is applied by using in the sensor two source-detector distances: 3 cm from the source and 4 cm from the source, which penetrates deeper into the brain. The SRS method leaves a signal originating predominantly in the brain cortex.

Oximeter provides a "predominately cerebral" measurement where over 85 percent of the signal, on average, is exclusively from the brain" (www.somanetics.com).

## 5. The HEG

The Toomim's hemoencephalograph (HEG) uses two wavelengths, 660 and 850 nm is a single-distance CW spectrophotometer. The distance between the source and receiver is 3 cm. The HEG measures the intensity ratios of the 660 nm light to 850 nm light. The HEG does not measure $rSO_2$, but is an important tool in the biofeedback research. The HEG is a very sensitive device. The source-detector is 3 cm, while the Somanetics INVOS Cerebral the oximeter has two distances 3 and 4 cm. Therefore, the INVOS Oximeter penetrates deeper into the brain and is more stable and less influenced by biofeedback.

A normalization of the HEG readings was done using measurements of 154 adult attendees at a professional society meeting, which is not directly related to rSO2. We have shown (Gersten et. al. 2011a) that one can relate the readings of the HEG to $rSO_2$ and even calibrate it separately for each individual. An improved derivation is given below.

## 6. Evaluating the $rSO_2$ from ratios of intensities

The regional hemoglobin saturation ($rSO_2$) is defined as

$$rSO_2 = 100 C_{HbO2} / \left( C_{HbO2} + C_{Hb} \right) \%, \qquad (7)$$

which is the percentage of the total hemoglobin that is oxygenated. Above $C_{Hb}$ and $C_{HbO2}$ are the concentrations of deoxyhemoglobin and oxyhemoglobin respectively. The modified Beer-Lambert law, Eq. (4), for a definite wavelength $\lambda$ is,

$$A(\lambda) = log_{10}\left(I_0(\lambda)/I(\lambda)\right) = \left[a_1(\lambda)\cdot c_1 + a_2(\lambda)\cdot c_2 + \ldots + a_n(\lambda)\cdot c_n\right]\cdot d\cdot DPF + G. \qquad (8)$$

It will be more convenient to work with natural logarithms $ln$;

$A(\lambda) = log_{10}(I_0(\lambda)/I(\lambda)) = -ln(I(\lambda)/I_0(\lambda))/00(10),$

Let us define the ratio of the relative intensities with wavelengths $\lambda_1$ and $\lambda_2$ as,

$R(\lambda_1, \lambda_2) = (I(\lambda_1)/I_0(\lambda_1))/ (I(\lambda_2)/I_0(\lambda_2)),$
$ln(R_{1,2}) = ln(R(\lambda_1, \lambda_2)) = ln(10)\times(A(\lambda_2)-A(\lambda_1)),$

The scattering component G is generally unknown and is highly dependent on the measurement geometry making it difficult to obtain absolute value of attenuation as a function of chromophore concentrations. However, under the assumption that G does not change during the measurement period, it is possible to determine changes in attenuation. Under this assumption, using Eq. (8), G is cancelled out in $A(\lambda_2)-A(\lambda_1)$, and we obtain,

$$ln(R_{1,2}) = ln(10)[(aHb((\lambda_2)) - a_{Hb}(\lambda_1))C_{Hb} + (a_{Hb\,O2}(\lambda_2) - a_{Hb\,O2}(\lambda_1))C_{Hb\,O2}]d\cdot DPF. \qquad (9)$$

Let us introduce the constant D:

$$D = ln(10)d\cdot DPF, \qquad (10)$$

Then

$$ln(R_{1,2}) = [(a_{Hb}(\lambda_2) - a_{Hb}(\lambda_1))(C_{Hb} + C_{Hb\,O2}) + (a_{Hb\,O2}(\lambda_2) - a_{Hb}(\lambda_2) - a_{Hb\,O2}(\lambda_1) + a_{Hb}(\lambda_1))C_{Hb\,O2}]D =$$
$$(C_{Hb} + C_{Hb\,O2})[100(a_{Hb}(\lambda_2) - a_{Hb}(\lambda_1)) + (a_{Hb\,O2}(\lambda_2) - a_{Hb}(\lambda_2) - a_{Hb\,O2}(\lambda_1) + a_{Hb}(\lambda_1))rSO_2]D/100. \qquad (11)$$

Let us introduce the following constants:

$$C_{Tot} = C_{Hb} + C_{Hb\,O2} \quad \text{the total hemoglobin,} \qquad (12)$$

$$H\left(\lambda_1, \lambda_2\right) = 100(a_{Hb}(\lambda_2) - a_{Hb}(\lambda_1)), \tag{13}$$

$$K\left(\lambda_1, \lambda_2\right) = a_{Hb\,O2}(\lambda_2) - a_{Hb}(\lambda_2) - a_{Hb\,O2}(\lambda_1) + a_{Hb}(\lambda_1), \tag{14}$$

$$M = C_{Tot} \cdot D/100, \tag{15}$$

Then,

$$ln\left(R_{1,2}\right) = M \cdot (H\left(\lambda_1, \lambda_2\right) + K\left(\lambda_1, \lambda_2\right) \cdot rSO_2). \tag{16}$$

Above, the constant M is not known. One can determine it in two ways, by introducing a third wavelength, or by calibrating with other device. If third wavelength is added one obtains,

$$ln\left(R_{1,2}\right)/ln\left(R_{1,3}\right) = (H\left(\lambda_1, \lambda_2\right) + K\left(\lambda_1, \lambda_2\right) \cdot rSO_2)/(H\left(\lambda_1, \lambda_3\right) + K\left(\lambda_1, \lambda_3\right) \cdot rSO_2), \tag{17}$$

from which $rSO_2$ can be evaluated. If one would like to calibrate M using other device, one can calibrate M only with respect to a standard total hemoglobin $C_0$,

$$M = (C_{Tot}/C_0) \times \left(C_0 \times D/100\right) = (C_{Tot}/C_0) \times M_0, \tag{18}$$

Where $C_{Tot}$ can be determined from a blood test.

## 7. Peculiarities of brain's blood flow: Role of carbon dioxide

Breathing may have dramatic effects on the brain blood flow. This was already known long time ago to Chinese, Indians and Tibetans. We will develop simple mathematical models which allow a quantitative description of cerebral blood flow. In recent years considerable progress was made in utilizing measurements of the regional cerebral blood flow (rCBF) in order to study brain functioning [44, 2, 8]. It seems however that the physical and mathematical aspects of the global cerebral blood flow (CBF), or average rCBF, were not sufficiently explored. Our main interest is the use of physical principles [35], physical and mathematical reasoning as well as means to describe the main features of CBF in a simple way.

The human brain consists of about 2% of the adult body weight, but consumes (at rest) about 15% of the cardiac output (CO) and about 20% of the body's oxygen demand [61, 25]. Glucose is the main source of cellular energy through its oxidation [61]. The cerebral glucose utilization is almost directly proportional to the CBF, [29; 33]. The CBF can be influenced by abnormal glucose levels, is increased during hypoglycemia [36] and decreased during hyperglycemia [10]. Normal mean CBF is approximately 50-55 ml/100g/min, but declines with age (above the age of about 30), in a rate of approximately $58.5-0.24 \times$age ml/100g/min [50, 27], see also [69], for other details).

The cardiac output can be increased many times (up to about tenfold) during very heavy exercise or work. Only part of the cardiac output increase can be accommodated by the brain blood vessels because of autoregulatory mechanisms and because of the vessels limited capacitance, which is influenced by their elasticity, limited space of the cranium and the presence of the cerebrospinal fluid (CSF).

Autoregulatory mechanisms exist, which maintain the CBF approximately constant for cerebral perfusion pressure (CPP) over an approximate range of 60-160 mm Hg [29,[3] 67]. Outside this autoregulatory range the CBF may decrease (CPP<60 mm Hg) as in the case of hypotonia [61] or increase (CPP>160 mm Hg) as in the case of high hypertension [25]. Again, the above statements are valid only for normal functioning. For some abnormal functioning the autoregulatory mechanisms may break down, for example if $PaCO_2$> 70 mm Hg [29, 30].

The CBF is also influenced by the value of cerebral tissue $PaO_2$, whose normal range is about 100 mmHg. Only below approximately 40-50 mmHg there will be a very strong increase of CBF [25], mobilizing the organism to prevent suffocation. The main parameter influencing the CBF is the arterial $PaCO_2$. About 70% increase (or even less) in arterial $PaCO_2$ may double the blood flow (normal value of $PaCO_2$ is about 40 mmHg.) [61, 25].

The CBF is very sensitive to $PaCO_2$and it is our aim to demonstrate with a simple physical model that important information about CBF capacitance can be obtained by considering only the dependence of CBF on $PaCO_2$. Slowing down the breathing rate, without enhancing the airflow [20, 18, 19], or holding the breath, can increase $PaCO_2$. It is plausible that this is one of the essences of yoga pranayama [4, 39, 40, 46], and of Tibetan six yogas of Naropa [13, 51]. It seems that biofeedback training of breathing [20], or methods advocated in yoga, may become important for treating health problems.

In Sec. 8 We have developed a simple physical model, and have derived a simple four parameter formula, relating the CBF to $PaCO_2$. With this model experimental data sets of rhesus monkeys and rats were well fitted. In Sec.9 exact formulae were found, which allow to transform the fits of one animal to the fits of another one. The merit of this transformation is that it allows to use rats data as monkeys data (and vice versa) simply by rescaling the $PaCO_2$ and the CBF data.

## 8. A mathematical model of CBF as a function of arterial $CO_2$

Inspection of experimental data, especially the more accurate ones on animals, like those done with rhesus monkeys [55], or with rats [59] led us to conclude that the CBF (which will be denoted later as B) is limited between two values. We interpreted this as follows: the upper limit $B_{max}$ corresponds to maximal dilation of the arterioles and the lower (non-negative) limit $B_{min}$ to the maximal constriction of the vessels. Reivich has fitted his data with a logistic model curve [55], which has two asymptotes

$$B(p) = \left( 20.9 + \frac{92.8}{1 + 10570 \exp[-5.251 \ \log_{10}(p/1mmHg)]} \right) ml / 100g / \min, \qquad (19)$$

where p= $PaCO_2$ in mmHg. Instead of the variable $B$ (the CBF) we will use the normalized to unity quantity $z$ defined as

$$z = \frac{B - B_{min}}{B_{max} - B_{min}}. \qquad (20)$$

The dependence of the CBF on $PaCO_2$ will be described with the dimensionless variable $x$=log($p/p_1$), where $p$= $PaCO_2$ in mmHg and $p_1$ is a fixed value of $PaCO_2$, which may be taken to be $p_1$=1mmHg. The variable $p$ is physical only for $p \geq 0$, in order to avoid formulae which may be valid for $p<0$ the variable $x$=log ($p/p_1$), valid for $p \geq 0$ was introduced. We can incorporate the above requirements and use the following assumptions:

$$\frac{dB}{dx} = AF(z) = AF\left( \frac{B - B_{min}}{B_{max} - B_{min}} \right), \qquad F(1/2) = 1,$$

$$0 \leq z \leq 1, \quad 0 \leq B_{min} \leq B \leq B_{max}, \qquad (21)$$

where A is a constant (reactivity), and $F(z)$ is a function which depends on CBF only. We will add the following boundary conditions on $F(z)$:

$$F(0) = F(1) = F'(0) = F'(1) = 0, \qquad (22)$$

where $F'(z) = dF(z)/dz$. The condition F(0)=0 corresponds to the requirement that the constriction is maximal at B=$B_{min}$, F(1)=0 correspond to maximal dilation for B=$B_{max}$. Another physical boundary constraint can be formulated for the derivatives $F'(z)$ as follows: $F'(0) = F'(1) = 0$, which means that the approach to the limits is smooth. We will assume that the constricting and dilating forces are the same; mathematically this condition can be expressed in the following manner,

$$F(\frac{1}{2}+z)=F(\frac{1}{2}-z), \quad or \quad F(z)=F(1-z). \tag{23}$$

There are many solutions, which satisfy the requirements Eqs. (21-23). We found the following ones:

$$F(z)=4^{n}z^{n}(1-z)^{n}, \quad and \quad F(z)=\sin^{n}(\pi z), \quad n \geq 2. \tag{24}$$

We will choose

$$F(z)=\sin^{2}(\pi z), \tag{25}$$

which will also enable us to integrate analytically Eq. (21) and to obtain a rather simple result to visualize. We will be able to utilize it for rescaling rats data to rhesus monkeys data and eventually to human data. This path will also enable us to translate rats data to monkey and eventually human data. From Eq.(20):

$$dB=\left(B_{\max}-B_{\min}\right)dz, \tag{26}$$

Eqs. (21) and (25) can be converted to

$$\frac{dz}{\sin^{2}(\pi z)}=\frac{Adx}{\left(B_{\max}-B_{\min}\right)}. \tag{27}$$

One can easily check that

$$\int\frac{dz}{\sin^{2}(\pi z)}=-\frac{1}{\pi}ctg(\pi z)+C, \tag{28}$$

where C is an arbitrary constant. From Eq. (28) we obtain,

$$\int_{z_1}^{z_2} \frac{dz'}{\sin^2(\pi z')} = -\frac{1}{\pi}ctg(\pi z_2) + \frac{1}{\pi}ctg(\pi z_1) = \frac{1}{\pi}\frac{\sin\pi(z_2 - z_1)}{\sin\pi z_2 \sin\pi z_1}. \tag{29}$$

Integrating Eq. (27), using Eq. (28), we obtain

$$\int_{z_r}^{z} \frac{dz'}{\sin^2(\pi z')} = -\frac{1}{\pi}ctg(\pi z) + \frac{1}{\pi}ctg(\pi z_r) = \int_{x_r}^{x} \frac{Adx}{\Delta B} = \frac{A(x - x_r)}{\Delta B} \tag{30}$$

were $\Delta B = B_{max} - B_{min}$. From Eq.(30) we obtain,

$$z = \frac{1}{\pi}arcctg\left(ctg(\pi z_r) - \pi\frac{A(x - x_r)}{\Delta B}\right) \tag{31}$$

From Eq. (31) and the relation: $arcctg(\pi z) + arctg(\pi z) = \frac{\pi}{2}$, we get:

$$z = \frac{1}{2} - \frac{1}{\pi}arctg\left(ctg(\pi z_r) - \pi\frac{A(x - x_r)}{\Delta B}\right) \tag{32}$$

Eq. (32) describes the dependence of CBF (z in Eq. (32)) on $p$ in terms of 5 parameters: $z_r$, $x_r$, $A$, $B_{min}$, $B_{max}$. The number of parameters can be further reduced to four.

### 8.1. The four parameter formula

A simple way to eliminate $z_r$ from Eq. (32) is to choose $z_r$=0.5 (the value half way between the extremes) then $ctg(\pi z_r) = 0$. Eq. (32) will have the simple form,

$$z = \frac{1}{2} + \frac{1}{\pi}arctg\left(\pi\frac{A(x - x_r)}{\Delta B}\right) \tag{33}$$

and after substituting $z = (B - B_{min})/\Delta B$,

$$B = B_0 + \frac{\Delta B}{\pi}arctg\left(\pi\frac{A(x - x_0)}{\Delta B}\right), \tag{34}$$

and after substituting $x=\log(p/p_1)$, the four parameter formula is

$$B(p) = B_0 + \frac{\Delta B}{\pi} arctg\left(\pi\frac{A\ln(p/p_0)}{\Delta B}\right), \tag{35}$$

where,

$$B_0 = B(z_r = \tfrac{1}{2}) = \tfrac{1}{2}(B_{max} + B_{min}); \quad \left(\frac{dB}{dx}\right)_{B=B_0}$$
$$= A, \quad x_0 = x(z_r = \tfrac{1}{2}), \quad p_0 = p(z_r = \tfrac{1}{2}). \tag{36}$$

## 8.2. Experimental data with rhesus monkeys and rats

To our knowledge the experimental data of Reivich [55] are the only published data which tabulate CBF and $PaCO_2$ for individual animals. This will enable us to check our model and find individual variations. The fit results are given in table 1. and displayed in Fig. 1. It should be noted that for monkey's No. 2 and 3 the data were insufficient to determine all the parameters. In Fig. 1, the dotted lines were obtain ned using the all monkeys parameters. This model seems to describe the main features of the data.

| Monkey No. | $B_{min}$ ml/100g/min | $B_{max}$ ml/100g/min | $p_0$ mmHg | A ml/100g/min |
|---|---|---|---|---|
| 1 | 44.7±5.2 | 95.2±4.8 | 61.0±2.1 | 147.7±76.6 |
| 2 | | not enough data | | |
| 3 | | not enough data | | |
| 4 | 17.6±2.6 | 125.4±3.3 | 91.4±5.0 | 82.2 ±12.2 |
| 5 | 14.7±2.5 | 123.1±2.8 | 54.6±2.0 | 95.6±11.9 |
| 6 | 9.2±3.1 | 121.1±3.0 | 70.6±4.7 | 48.2±5.6 |
| 7 | 18.3±1.3 | 86.9±2.0 | 43.9±1.7 | 84.7±8.8 |
| 8 | 27.3±2.7 | 142.2±3.7 | 82.7±1.3 | 416±145 |
| all monkeys | 13.6±2.5 | 122.5±2.7 | 60.3±1.9 | 79.3±8.2 |
| rats | 89.0±10.4 | 512.4±18.1 | 38.0±1.1 | 581.7±91.0 |
| all men | 23.3±1.9 | 165.9±3.9 | 53.2±0.7 | 251.5±18.1 |

**Table 1.** The fit of Eq. (36) to experimental data of individual monkeys [55], all off the monkeys and rats [59].

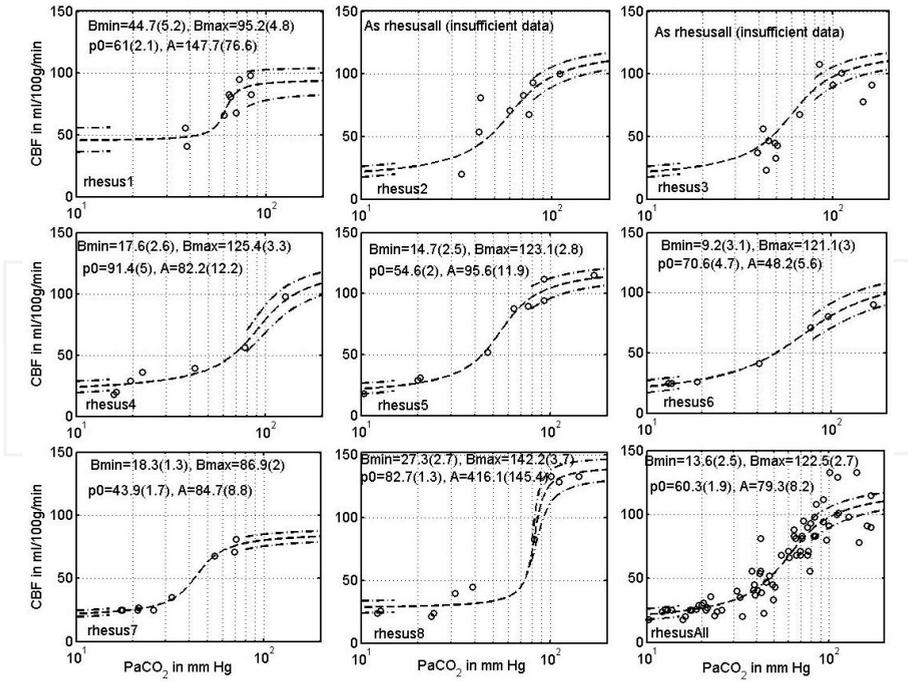

**Figure 2.** The fits to the experimental data of [55] with parameters given in Table 1. The dotted line is the rescaling of the monkeys best fit according to Eqs. (44) and (45).

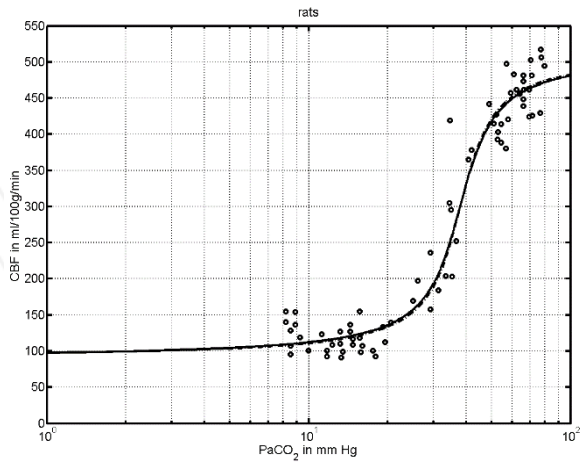

**Figure 3.** The dependence of CBF of rats on the partial tension of $CO_2$. The continuous curve is the best fit to Eq. (36) for the data of [59]. The parameters are given in Table 1.

## 9. Rescaling the data

In this section, we will transform rat data to serve as monkey data and vice versa. We will utilize Eq. (21), and use the upper index R to denote rats and upper index M to denote monkeys. Let us consider two separate fits (e.g. for all monkeys and for rats):

$$B^R(p^R) = B_0^R + \frac{\Delta B^R}{\pi} arctg\left( \pi \frac{A^R \ln(p^R / p_0^R)}{\Delta B^R} \right),$$

(37)

$$B^M(p^M) = B_0^M + \frac{\Delta B^M}{\pi} arctg\left( \pi \frac{A^M \ln(p^M / p_0^M)}{\Delta B^M} \right).$$

(38)

Let us assume the possibility that the curve (37) is converted to curve (38) and vice versa. We will show that it is possible to do this by rescaling the variables p (PaCO2) and B (CBF). Instead of demanding that Eq.(37) be equal to Eq. (38) we will require the equivalent conditions,

$$\pi \frac{A^R \ln(p^R / p_0^R)}{\Delta B^R} = \pi \frac{A^M \ln(p^M / p_0^M)}{\Delta B^M},$$

(39)

$$\frac{B^R(p^R) - B_0^R}{B^M(p^M) - B_0^M} = \frac{\Delta B^R}{\Delta B^M}.$$

(40)

Equation (39) transforms the p coordinates, and Eq.(40) transforms the B variables. Thus, for example, if we would like to transfer the monkey data to rat data we will rewrite Eq. (40) and Eq. (41) as follows,

$$p^R = p_0^R \exp\left( \frac{\Delta B^R A^M \ln(p^M / p_0^M)}{\Delta B^M A^R} \right),$$

(41)

$$B^R(p^R) = \frac{\Delta B^R}{\Delta B^M}\left( B^M(p^M) - B_0^M \right) + B_0^R.$$

(42)

Eq. (41) transforms the $p^M$ coordinates (using the parameters of Table 1.) into $p^R$ coordinates and Eq.(42) transforms the CBF $B^M$ data into $B^R$ data. The transition of the rat's data to monkey's data is achieved via the equations

$$p^M = p_0^M \exp\left(\frac{\Delta B^M A^R \ln(p^R / p_0^R)}{\Delta B^R A^M}\right), \tag{43}$$

$$B^M(p^M) = \frac{\Delta B^M}{\Delta B^R}\left(B^R(p^R) - B_0^R\right) + B_0^M. \tag{44}$$

In Fig. 3. The data of the rhesus monkeys (circles) with the fit of all monkeys (the line as in Fig. 1.) are supplemented with the rat data (stars) according to Eqs. (43) and (44).

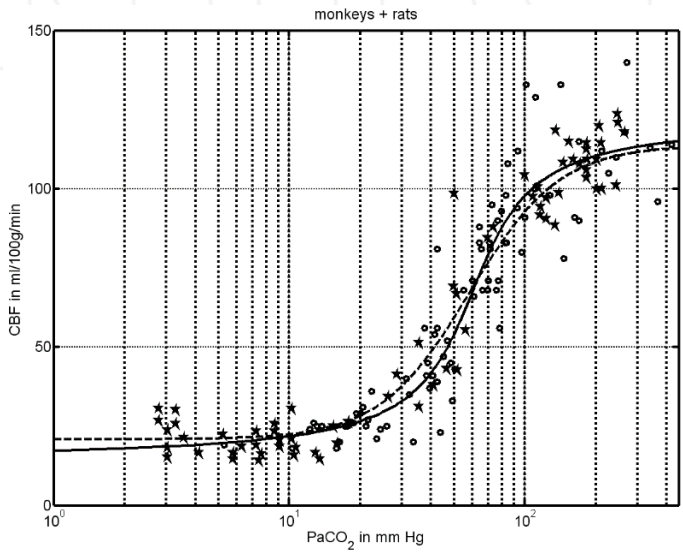

**Figure 4.** The data of the rhesus monkeys (circles) with the fit of all monkeys Eq.(46) (solid line) and Eq. (19) (dashed line) supplemented with the rat data (stars) according to Eqs. (43) and (44).

## 10. Human data

For obvious reasons there are no experimental data on individual humans available in a very wide range of $PaCO_2$. The measurements on animals usually ended with weighing their brains and calibrating the results for 100g of brain tissue. Recently restrictions were imposed on experiments in which animals are killed. In recent years, the emphasis was placed on getting regional CBF (rCBF) measurements rather than global CBF. As a result, the extended animal measurements are rather old.

Our fit to human data, based on (Reivich 1964, Ketty and Schmidt 1948 and Raichle et all 1970) is given in Fig. 4, with parameters of Table 1 (all men). In [55] it was observed that the human data (in a narrow interval of PaCO2), which existing in that time, were within the experimental errors very close to the rhesus monkey data. One can see it in Fig. 4 (circles, the data of [55]. Therefore the fit of [55], Eq. (19) of the rhesus monkey data was and is still being used as the relation between $PaCO_2$ and CBF of humans.

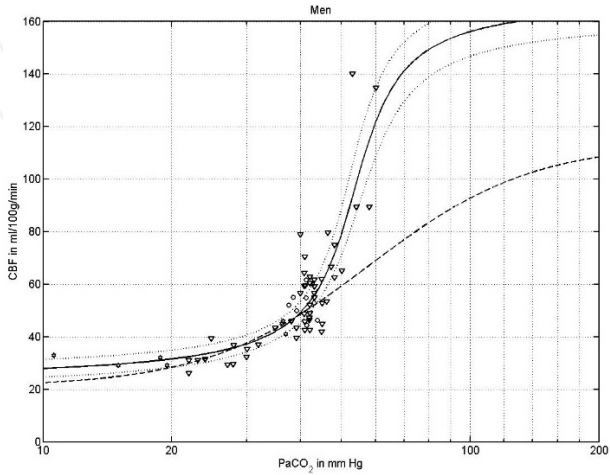

**Figure 5.** Human CBF data [55], circles), [42], triangles) and [53], hexagons). The dashed line of the fit of Eq. (19) represents the rhesus monkey data. The dotted lines deviate from the solid line, given by Eq. (46), by shifting all parameters by two error bars.

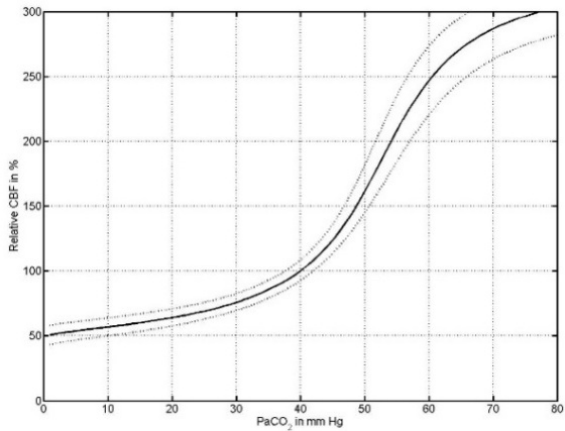

**Figure 6.** Relative changes of human CBF with respect to its values at $PaCO_2$=40 mm Hg.

Our fit to the rhesus monkey data (from Table 1.) is,

$$B(p) = 68.0 + 34.7\, arctg\left(2.29\ln(p/60.3mmHg)\right) ml/100g/\min \tag{45}$$

The fits of Eqs. (10.1) and (10.2) are quite similar and are depicted in Fig. 3. They may serve as a first estimate of human CBF below $PaCO_2$=45 mm Hg. Eq. (10.1) represents the human data in some medical textbooks and publications without mentioning that it is a fit of Rhesus monkeys (for example: [25]. Our fit to human data

$$B(p) = 94.6 + 45.4\ arctg\left(5.54\ln(p/53.2mmHg)\right) ml/100g/\min \tag{46}$$

deviates from that of the Rhesus monkeys in the hypercapnia region ($PaCO_2$ above 45 mm Hg). More accurate human data in this region are needed to confirm our fit. Changes in $PaCO_2$ induce dramatic changes in CBF. This can be seen in Fig. 5, where relative changes of CBF are given. One can see that changing $PaCO_2$ from normal (40 mm Hg) by 10 mm Hg can change CBF by about 25%-55%.

## 11. Experimental detection of oxygen waveforms

In (Gersten et al., 2009) we have presented simultaneous measurements of EtCO2 (by capnometer, closely related to PaCO2) and rSO2 (with the INVOS Oximeter of Somanetics model 5100B) during intensive breathing exercises.

The fastest recording rate of the 5100B oximeter was 12. This time is much longer than the average period, of about 4 seconds, of normal respiration. In order to detect oxygenation periodicity we had to study respiration periods of about 36 seconds or larger (i.e. 3 data points or more for each breathing period). This is still a rather small amount of data points per respiration period. We have compensated for this small number by using a cubic spline interpolation of the data points, adding new interpolation points through this method. The cubic spline interpolation is a very effective method of smooth interpolation. We found six people well acquainted with yoga pranayama, who could easily perform breathing exercises with periods around 36 seconds. All of them performed the following routine which lasted for 15 minutes. They were asked to breathe in the following way: to inhale for 4 units of time (UOT), to hold the breath for 16 UOT and to exhale for 8 UOT, this we denote as the 4:16:8 (pranayama) routine. The unit of time (UOT) is about 1 second. The yoga practitioners develop an internal feeling of UOT which they employ in their practices. They learn to feel their pulse or they learn to count in a constant pace. Often they practice with eyes closed. In order not to distract or induce additional stress we preferred not to supply an external uniform UOT. The primary concern for this research was to have a constant periodicity and in this case the practitioners have succeeded to maintain it. The data were analyzed with spectral analysis which took into account non-stationary developments, which were subtracted from $rSO_2$ and BVI data. Sharp picks corresponding to the breathing periodicity were found in the spectral analysis of $EtCO_2$ (the amount of $CO_2$ during expiration), the overlapping of the periodicities is shown in Fig. 7 (subfigures 2a, 2b, 2c, 2d, 2e, 2f) and Table 2.

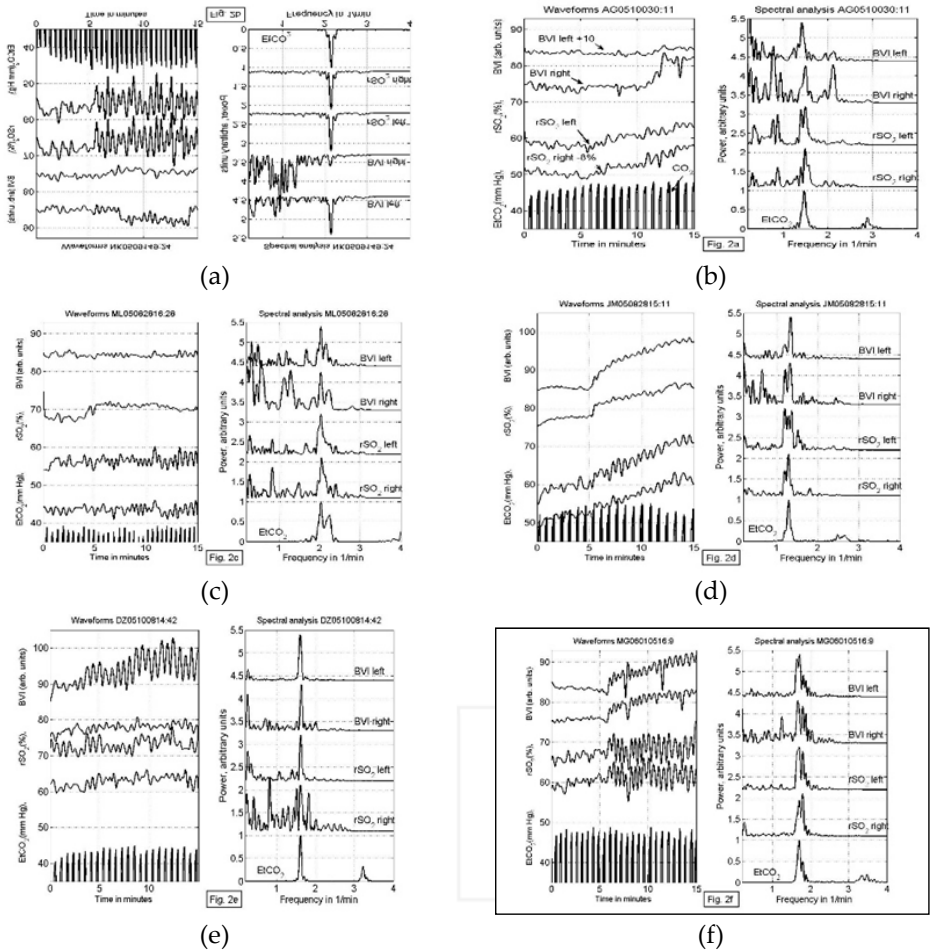

(a)

(b)

(c)

(d)

(e)

(f)

**Figure 7.** subfigures 2a-2f). On the perpendicular axis are the readings of the capnometer, $rSO_2$ right sensor (subtracted with 8%), $rSO_2$ left sensor, BVI right sensor, BVI left sensor (increased by 10). On the right hand side the corresponding spectral analyses of the waveforms are given, from [22].

| | $PaCO_2$ | $rSO_2$ right | $rSO_2$ left | BVI right | BVI left |
|---|---|---|---|---|---|
| **AG** | 1.47 (1.42-1.51) | 1.49 (1.45-1.53) | 1.48 (1.45-1.54) | 1.49 (1.42-1.53) | 1.42 (1.38-1.47) |
| **NK** | 2.15 (2.12-2.18) | 2.14 (2.12-2.17) | 2.14 (2.12-2.17) | ------------------ | 2.14 (2.11-2.17) |
| **ML** | 2.03 (1.98-2.07) | 2.05 (2.01-2.15) | 2.03 (1.95-2.08) | 2.03 (1.99-2.07) | 2.03 (2.00-2.06) |
| **JM** | 1.30 (1.26-1.34) | 1.30 (1.26-1.33) | 1.30 (1.17-1.40) | 1.33 (1.29-1.37) | 1.35 (1.31-1.38) |
| **DZ** | 1.62 (1.59-1.65) | 1.62 (1.59-1.67) | 1.62 (1.60-1.65) | 1.63 (1.61-1.67) | 1.61 (1.57-1.65) |
| **DG** | 1.70 (1.65-1.73) | 1.79 (1.65-1.82) | 1.70 (1.62-1.82) | 1.66 (1.62-1.89) | 1.71 (1.61-1.74) |

**Table 2.** The position of the dominant frequencies (in units of 1/minute) of Fig. 7 in the spectral analysis, in parenthesis the extension of the half width is given, from [22]

## 12. Simple exercises

Significant increase of $PaCO_2$ (and of total CBF) can be achieved with untrained people using very simple breathing procedures. The reason for that is the dependence of $PaCO_2$ on ventilation [68],

$$PaCO_2 = K \, (\dot{V}_{CO_2} / \dot{V}_A), K = 863 \text{ mm Hg,} \qquad (47)$$

where $\dot{V}_{CO_2}$ is the metabolic $CO_2$ production and $\dot{V}_A$ is the alveolar ventilation.

In [22] we have described the influence on the brain oxygenation of simple exercises performed by 18 students for the first time. The exercises, each lasting for 5 minutes, were:

1. Simple breathing exercises. It was possible to increase the $PaCO_2$ by either breathing slowly or by holding the breath.

2. Simple arithmetic counting, concentrating on a mental problem changes brain's oxygenation locally [5].

3. Biofeedback, as observed by Hershel [66]

In [23] we calibrated the HEG readings using the $rSO_2$ readings of the INVOS oximeter of Somanetics. We found,

$$x_1/x_2 = ln(y_1/32.08) / ln(y_2/32.08), x \equiv rSO_2, y \equiv \text{HEG readings.} \qquad (48)$$

Simultaneous measurements were taken using HEG and a capnometer. Eighteen students participated in the experiment in which HEG and $CO_2$ data were recorded for 5 intervals of baseline, simple breathing exercises, simple arithmetic tasks and biofeedback. The results show that almost all participants could increase their brain oxygenation or CBF, but in each case it was strongly dependent on one of the three methods used. We can conclude that it is possible to substantially increase local oxygenation or global CBF using one of the three methods described above, but the preferred method is highly individual. The protocol of this research was approved by the IRB of Hunter College of the City University of New York.

The participants were 18 participants from the introductory course to psychology (PSY 100) in Hunter College of the City University of New York. All participants had to sign an informed consent. At least two experimenters were present during each experiment. The confidentiality of the participants was protected. Illustrative examples from [22] are given below.

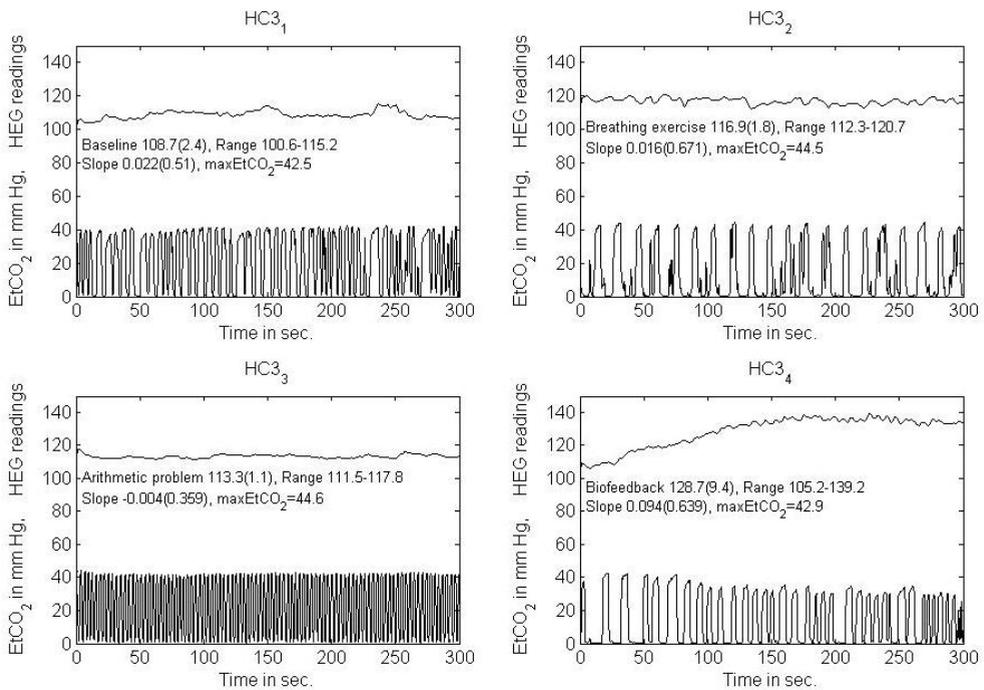

**Figure 8.** The HEG and the capnometer readings of participant No. 3. SD in parenthesis [22].

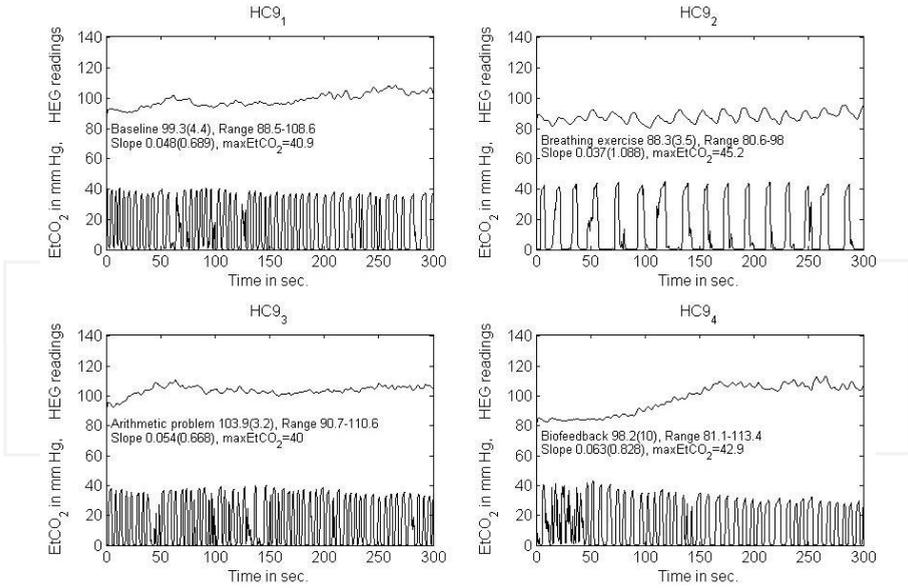

**Figure 9.** The HEG and the capnometer readings of participant No. 9. SD in parenthesis [22].

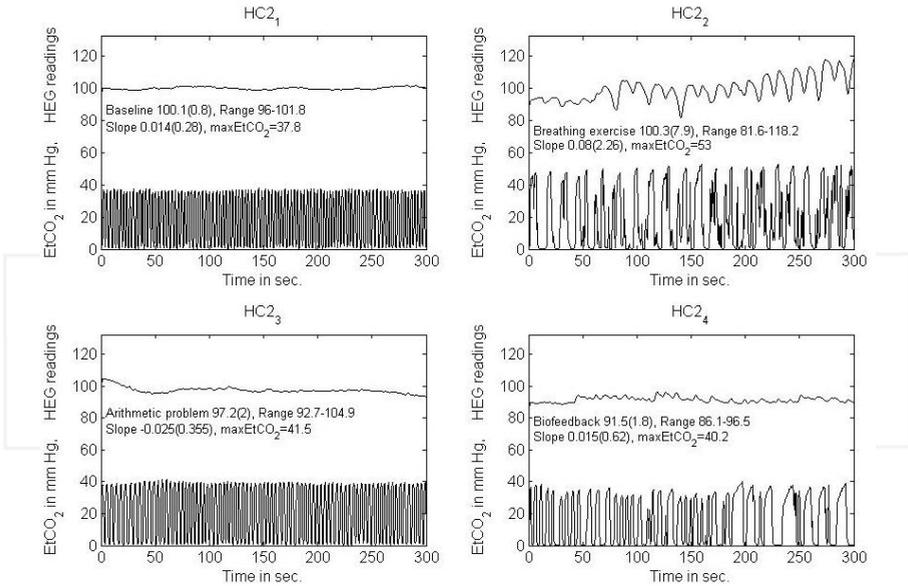

**Figure 10.** The HEG and the capnometer readings of participant No. 2. SD in parenthesis [22].

In most cases the HEG and the $CO_2$ waveforms had the same periodicity, which is well depicted in Fig. 11.

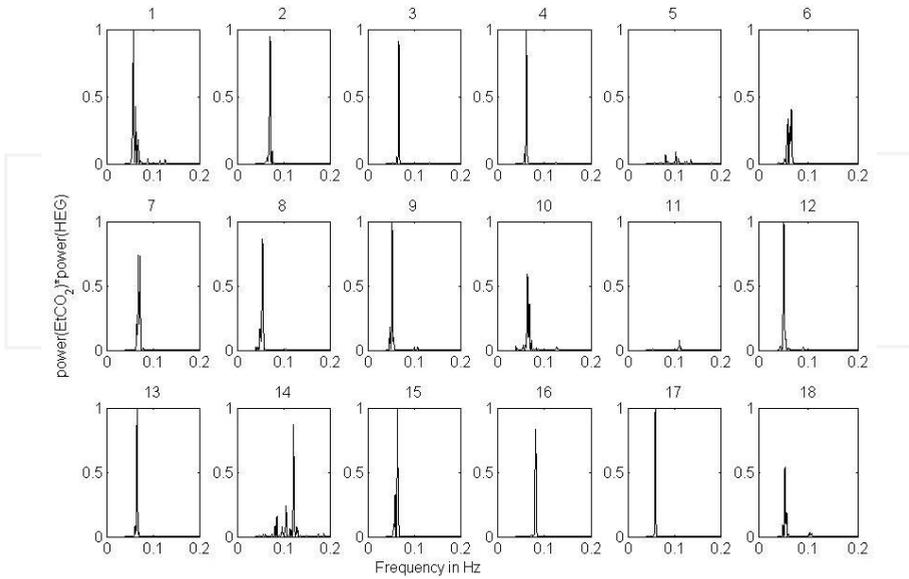

**Figure 11.** The correlation between the power spectra of the $EtCO_2$ periodic pattern and the corresponding HEG periodic pattern is depicted by their multiplication. The power spectra are normalized to unity. Maximal correlation is achieved when the multiplication is equal to one [22].

Below, in Table 3, one can observe the remarkable increase of PaCO2 with very simple breathing exercises. One should remember that this was a first attempt, without previous training.

| N | Baseline | Breath. Ex. | Arith. Prob. | Biofeedback | % increase |
|---|----------|-------------|--------------|-------------|------------|
| 1 | 37.5 | 46.9 | 39.4 | 34.5 | 25.2 |
| 2 | 36.6 | 44.9 | 39.2 | 34.4 | 22.8 |
| 3 | 40.1 | 40.7 | 42.1 | 34.0 | 1.5 |
| 4 | 35.4 | 38.4 | 38.0 | 32.0 | 8.3 |
| 5 | 37.5 | 36.7 | 38.5 | 39.1 | -2.0 |
| 6 | 36.9 | 39.6 | 35.6 | 36.3 | 7.4 |
| 7 | 31.4 | 39.8 | 38.3 | 38.3 | 27.0 |
| 8 | 37.5 | 46.2 | 40.2 | 38.5 | 23.2 |
| 9 | 37.5 | 41.9 | 35.5 | 34.6 | 11.7 |

| N | Baseline | Breath. Ex. | Arith. Prob. | Biofeedback | % increase |
|---|----------|-------------|--------------|-------------|------------|
| 10 | 33.3 | 38.8 | 36.4 | 36.4 | 16.5 |
| 11 | 37.1 | 39.6 | 37.5 | 38.6 | 6.5 |
| 12 | 40.0 | 43.7 | 40.3 | 40.6 | 9.1 |
| 13 | 38.2 | 45.9 | 39.3 | 38.7 | 20.1 |
| 14 | 33.3 | 34.1 | 35.9 | 36.3 | 2.4 |
| 15 | 38.2 | 45.7 | 42.3 | 40.6 | 19.8 |
| 16 | 28.6 | 33.9 | 28.9 | 28.8 | 18.4 |
| 17 | 40.3 | 43.3 | 41.2 | 39.0 | 7.6 |
| 18 | 35.4 | 42.6 | 40.6 | 37.9 | 20.3 |
| mean(SD) | 36 (3) | 41.3(4.0) | 38.3(3.2) | 36.6(3.1) | 13.7(8.8) |

**Table 3.** Mean values of $EtCO_2$ in mm Hg. Last column is the increase in % due to breathing exercises compared to baseline.

## 13. Autoregulation

The autoregulation of cerebral blood flow (CBF), or the independence of CBF on changes of mean arterial blood pressure (MABP) in a wide range of MABP (the so called "plateau"), is considered to be a well established fact. But looking carefully at the existing experimental data we could not find even one publication which gives a proper experimental support for the existence of the plateau of the autoregulation.

The first publication, seemingly proving autoregulation in humans, was that of [48]. However, the data at different points on the plateau were taken from different people [48], Reivich 1969). As individual differences are quite important, the above procedure is not accurate and may serve as an indication only. [30] has done experiments with 12 dogs, measuring CBF for a wide range of MABP. However, he presented and discussed results taken together on 8 normocapnic animals and 4 hypercapnic ones. Fortunately, he tabulated the experimental data for each dog separately. Harper's data are the only published data, which describe for each animal separately the dependence of the CBF on MABP in a wide range of MABP. We shall use extensively Harper's data and analyze their content.

[49] have done measurements of CBF on baboons for a wide range of MABP. Unfortunately the data were analyzed collectively i.e. on several baboons simultaneously. Individual baboons were not tabulated. Averaged data are supporting autoregulation. As one can see from this introduction, there is no firm experimental basis for autoregulation. In the following sections, we shall propose simple models that go beyond autoregulation but contain its main elements. For more recent reviews of the experimental and theoretical situation, the reader is encouraged to consult the reviews of [3, 24, 62, 67] and the references therein. The book of [26] has complementary material and references.

### 13.1. A generalization of the simple model of autoregulation

The oversimplified classical picture of autoregulation is depicted in Fig. 12. In this picture for MABP's in the range of 60-130 mm Hg there is no change in the CBF. This range may change somewhat for the cases of hypertension and hypotension. The plateau is shifted to the right in the case of hypertension, and to the left in the case of hypotension. Let us explain the picture in Fig. 12 in terms of a simple control feedback model. Without any feedback the CBF is expected to be proportional to the MABP

$$CBF(MABP) = S \cdot MABP, \tag{49}$$

where S is a constant describing the sloap of the to the flow in units of ml/min/100g/mmHg. In order to get the plateau the line of Eq.(13.1), the CBF should be diminished from the values of Eq.1 by a contribution, which is different for various MABP's. In a feedback control model this subtracted quantity is -F(MABP) CBF, where F(MABP) is a feedback (gain) function depending on MABP. Eq.(13.) should be now replaced with

$$CBF(MABP) = S \cdot MABP - F(MABP) \cdot CBF(MABP) . \tag{50}$$

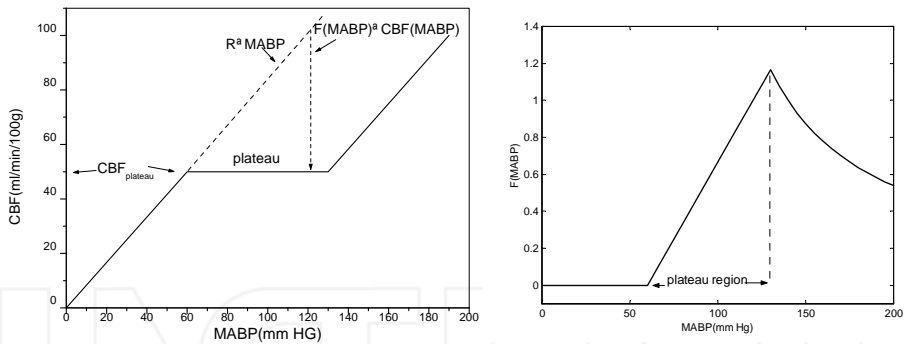

**Figure 12.** The classical picture of autoregulation. For MABP's in the range of 60-130 mm Hg there is no change in the CBF. The feedback function F of Eq.(53) as a function of MABP. In the plateau region it is linear.

By bringing the second l.h.s. term of Eq.(50) to the left we obtain

$$CBF(MABP) + F(MABP) \cdot CBF(MABP)= (1+ F(MABP)) \cdot CBF(MABP)= S \cdot MABP$$

and after dividing both sides by 1+ F(MABP) we finally obtain

$$CBF(MABP) = \frac{S \cdot MABP}{1+F(MABP)}. \tag{51}$$

Eq.(47) can also be written in the following form

$$F(MABP) = \frac{S \cdot MABP}{CBF(MABP)} - 1. \tag{52}$$

In many instances the constant S may be determined relatively well from experiments in which CBF is measured at different MABP's. In these cases Eq.(52) is very advantageous, as it allows to determine F(MABP) directly from the experiments. We shall extensively employ this procedure analyzing [30] experiments.

In order to get the continuous line and the plateau of Fig.12, the feedback function F(MABP) has to have the following form

$$F(MABP) = \begin{cases} 0, & for \quad 0 \le MABP \le 60 mmHg \\ \dfrac{R \cdot MABP}{CBF_{plateau}} - 1, & for \quad 60 mmHg \le MABP \le 130 mmHg, \\ \dfrac{MABP}{MABP - 70 mmHg} - 1, & for \quad MABP \ge 130 mmHg \end{cases} \tag{53}$$

where $CBF_{plateau}$ is the value of CBF on the plateau (here 50 ml/min/100g, as an example). It should be stated that the behavior of the CBF above 130 mm Hg is only a guess and is uncertain. We see a sharp raising of the CBF in the data of [49], but we do not se this raising in the data of Harper. For reasons of symmetry we have assumed the same slope above MABP of 130 mm Hg as below 60 mmHg. That is the reason why we have pointed out in Fig.12 that the models above MABP equal to 130 mm Hg are uncertain. The dependence of F, as given by Eq.(53) on MABP is depicted in Fig.12. One should note that in the plateau region the dependence on MABP is linear and thus very simple. We shall generalize this model by allowing linear dependence of F on MABP, but with different strengths. Above MABP of 130 mm Hg one may see in Fig.12 the breakdown of the linearly growing biofeedback contribution, but one should remember that our knowledge of what is going on in this region is quite uncertain.

The autoregulation as being depicted in Fig.12 is an over idealized picture. We know that very strong deviations from this picture exists in the case of hypercapnia, for which there is almost no feedback suppression as well as autoregulation. We shall consider such effects by changing the feedback of Eq. (50) by

$$CBF(MABP) = R \cdot MABP - c \cdot F(MABP) \cdot CBF(MABP), \tag{54}$$

i.e. we have changed the overall strength of the feedback by multiplying the feedback function by a constant which we denote as c. Eq.(53) is obtained with c=1, i.e. there is no change. In Fig.

13 we consider the three cases: c=1, c=0.1, 1.3. The c=0.1 case represents a strong suppression of the feedback, as we shall see later on this is typical for hypercapnia.

The case of c=1.3 represents an increase in feedback. As a consequence the CBF is supressed and is decreasing in the plateau region with the increase of MABP. We shall find such a behavior in some of the dogs of [30]. To our best knowledge this a new effect not described in the literature. We suspect, extrapolating from hypercapnia (for which c<<1), through normocapnia (for which autoregulation is assumed, i.e. c~1) to hypocapnia, that in the last case c>1.

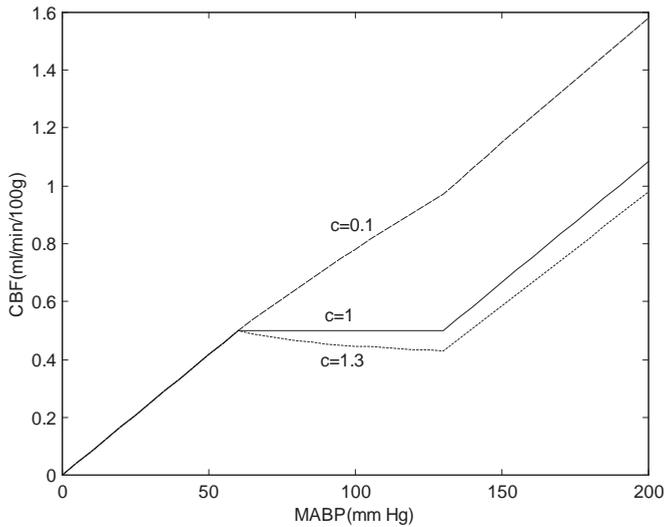

**Figure 13.** The change in CBF as a result of the change in the feedback strength. The constant c is a multiplication factor which multiplies the feedback function F(MABP).

## 13.2. Harper's experiments with dogs

[30] performed an experiment with 12 dogs in which he measured CBF while changing MABP in a wide range. His paper contains tables of the experimental data for each dog separately. To our best knowledge, these are the only published data of individual animals for which CBF was measured in a wide range of MABP. As we shall see later on, the parameters describing CBF in terms of MABP differ to great extent between the animals. Therefore, by trying to make a better statistics incorrect conclusions may be derived. For example, Harper in his paper claims: "Over a fairly wide range of blood pressure (from 90 to 180 mm Hg) the blood flow remained relatively constant, despite a varying blood pressure. This phenomenon will hereafter be referred to as 'autoregulation'." For some averaging procedures, this may look so, but as we shall analyze each dog separately, we shall notice rather large deviations. In our analysis we shall use the simple model of autoregulation Eq.(53) modified by Eq.(54). In the

data of the dogs, we do not see the sharp raise of CBF at larger MABP's; therefore we shall use the following model of the feedback function

$$F(MABP) = \begin{cases} 0, & for \quad 0 \le MABP \le MABP_1 \\ c \cdot \left( \dfrac{R \cdot MABP}{CBF(MABP_1)} - 1 \right), & for \quad MABP_1 \le MABP \end{cases}, \tag{55}$$

where the constant c was introduced to indicate the change with respect to the ideal autoregulation, i.e. the case where the CBF do not change with MABP in the plateau region. For ideal autoregulation c=1. The above model of the feedback assumes that the feedback function remains linear; the difference with the ideal autoregulation is only in the strength of the feedback.

The model has three parameters R, $MABP_1$, c. One of them is the threshold arterial mean blood pressure $MABP_1$ below which there is no feedback. Below $MABP_1$ the CBF depends only on the slope parameter R, CBF=R MABP. At the transition point $MABP_1$

$$CBF(MABP_1) = R \cdot MABP_1. \tag{56}$$

Substituting Eq.(56) into Eq.(55) we obtain

$$F(MABP) = \begin{cases} 0, & for \quad 0 \le MABP \le MABP_1 \\ c \cdot \left( \dfrac{MABP}{MABP_1} - 1 \right) & for \quad MABP \ge MABP_1 \end{cases}, \tag{57}$$

while CBF is obtained through Eq.(54).

The parameters were determined in the following way. First the parameter R was determined from the line starting from zero and tangential to the experimental points (not intersecting the lines connecting experimental points). Next, using Eq.(52) the experimental values of F(MABP) were determined from the value of R and experimental values of CBF(MABP). The feedback function was obtained by fiting the experimental data with Eq.(57), a linear fit. The parameters so determined for all the 12 dogs separately are given in table 1. To the table were also added the avarage values of $PaCO_2$ for each dog. For dogs B9-B12, which were in hypercapnia, there was practically no autoregulation, and the CBF data could be fitted with strait lines.

One should note that in Table 4 the values of the parameter c are far away from the value c=1, which correspond to the ideal autoregulation. One should also note the two cases (Dogs B6 and B7) for which c>1, which differ from previous cases in that their CBF may go down with increasing MABP

Table 4. The parameters of fitting the CBF equations.

Above, we have analyzed the data obtained by [30] in 12 Dogs, in which CBF was measured over a wide range of MABP. This is, to our knowledge, the only publication in which the data for each individual animal were tabulated. Contrary to the belief that these data support the picture of classical autoregulation, i.e. that CBF is almost constant in the plateau region, we found a somewhat different picture. The analysis of the data for each animal separately indicates that large deviations from the classical autoregulation may exist. We were able to interpret these data by a simple model that is based on the following assumptions: 1) up to a threshold level of MABP, denoted as *MABP1*, CBF is directly proportional to MABP (as in a rigid pipe). Above *MABP1* up to a level of MABP2, at which breakthrough occurs, there is a regulated suppression of CBF which can be explained by a negative feedback on CBF. This feedback is well described by a linear function of MABP (see Eq.(57) with a slope proportional to the parameter $\beta$ which may vary considerably among different individuals. The classical autoregulation model with a plateau between *MABP1* and *MABP2* is a particular case of this model with $\beta=1$. This model describes quite well the results obtained in dogs (Harper 1966) for which, as seen in Fig. 14, the individual feedback slope parameter varied to great extent, indicating the importance of using data obtained in individuals rather than the averaged data obtained for different individuals. Blood pressure medications are prescribed with the assumption of ideal autoregulation. Many side effects of these medications may result from disturbing the CBF.

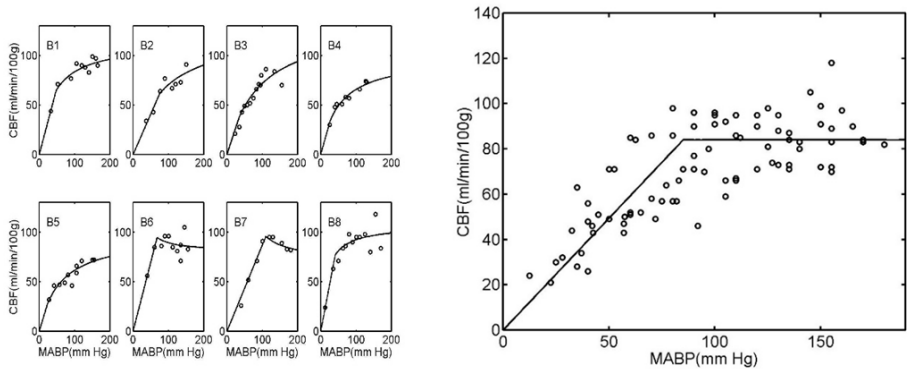

**Figure 14.** On the left individual expose a large deviation from ideal autoregulation. On the right all data do not exclude an ideal autoregulation.

## 14. Conclusions

Probing brain oxygenation is an important subject in medicine, physiology, psychology and education. Although the near infrared spectroscopy (NIRS) is not as powerful as MRI or fMRI, it is more available, less expensive, easier to handle, non-invasive and can be done in quite different circumstances, like in different positions or motion. In our experiments, we have used INVOS Cerebral Oximeter (ICO) and the spectrophotometer HEG described in Secs 4 and 5.

Breathing affects to great extent the brain physiology. Arterial partial pressure of carbon dioxide ($PaCO_2$) is among the major factors controlling the cerebral blood flow (CBF). The results can change dramatically even within a range of 50% from normal. The effect of $PaCO_2$ is peculiar in being almost independent of autoregulatory CBF mechanisms and allows exploring the full range of the CBF. In Sec. 8 a simple model and a simple formula (Eq. (35)), describing the dependence of the CBF on $PaCO_2$ was derived. Our model gives good quantitative predictions. Moreover, it allows imitating human data using animal data. With the results of our model, it is possible to devise breathing exercises and procedures which aim is to improve brains blood circulation.

The correlation between the $CO_2$ levels and the regional oxygen saturation ($rSO_2$) were studied in an experiment using very slow breathing patterns with the INVOS Cerebral Oximeter (ICO), with an important result of a definite periodic correlation between respiration, oxygenation and blood volume changes. The results shown in Sec. 11 are quite impressive. The deficiency of using the ICO was its very low sampling rate 1/12 Hz. For experimentation, we had to use subjects proficient in Yoga breathing.

The HEG had the advantage of a higher sampling rate of 1 Hz. With the HEG we could conduct experiments with untrained normal subjects. Our experiments with 18 student subjects is described in Sec. 12. Our three methods used in simple exercises can be used on the general population, are non-invasive, without the use of pharmaceuticals and have no side effects. They differ from each other in that the breathing affects mostly the global blood flow, arithmetic problem solving and biofeedback affects the regional blood flow (in our case the Fp1 region). Both our theoretical and experimental work differs from other studies due the specific instrumentation and our experimental procedure. Most of the results came close to our expectation.

We concluded that breathing can be used effectively to control CBF by the ventilatory control of end tidal $CO_2$. This research may have implications for complementary diagnosis and treatment of conditions involving regional cerebral metabolism such as cerebral vascular ischemia, seizures disorders, stroke, Alzheimer's disease, and more. Following that thought could lead us to improved cognitive function through a higher supply of oxygen to specific regions of the brain. We foresee future more detailed investigations to be made in the area of the effect of $CO_2$ on specific regions of the brain. This would be of great interest because a higher $CO_2$ supply results in a higher blood flow and thus to more oxygen and better overall brain function, specifically cognitive function.

As far as autoregulation is concerned, the individual data show a large deviation from the plateau model of autoregulation. Based on individual differences, a statistics based model may lead to incorrect administration of medications, which may influence strongly the cerebral blood flow and probably the blood flow to other organs. This may be the primary reason for side effects. Here experimentation with NIRS oximeters can be of great help for understanding how to treat individuals.

## Author details

Alexander Gersten[*]

Department of Physics, Ben-Gurion University of the Negev, Beer-Sheva, Israel